# Magnetism and exchange interaction of small rare-earth clusters; Tb as a representative

Lars Peters[1*], Saurabh Ghosh[2*], Biplab Sanyal[3†], Chris van Dijk[4], John Bowlan[5], Walt de Heer[6], Anna Delin[3,7], Igor Di Marco[3], Olle Eriksson[3], Mikhail I. Katsnelson[1,9], Börje Johansson[3] and Andrei Kirilyuk[1‡]

[1]Radboud University Nijmegen, Institute for Molecules and Materials, Heyendaalseweg 135, 6525 AJ Nijmegen, The Netherlands

[2] School of Applied and Engineering Physics, Cornell University, Ithaca, New York 14853, USA

[3]Department of Physics and Astronomy, Uppsala University, Box 516, 751 20 Uppsala, Sweden

[4]Van 't Hoff Institute for Molecular Sciences, University of Amsterdam, Science Park 904, 1098 XH Amsterdam, The Netherlands

[5]Fritz-Haber-Institut der Max-Planck-Gesellschaft, Faradayweg 4-6, D-14195 Berlin, Germany

[6]School of Physics, Georgia Institute of Technology, 837 State Str., Atlanta, Georgia 30332, USA

[7]Department of Materials and Nanophysics, School of Information and Communication Technology, Electrum 229, Royal Institute of Technology (KTH), SE-16440 Kista, Sweden

[8]SeRC (Swedish e-Science Research Center), KTH, SE-10044 Stockholm, Sweden

[9]Department of Theoretical Physics and Applied Mathematics, Ural Federal University, 620002 Ekaterinburg, Russia

†e-mail: biplab.sanyal@physics.uu.se
‡e-mail: a.kirilyuk@science.ru.nl
* These authors contributed equally to this work.

Here we follow, both experimentally and theoretically, the development of magnetism in Tb clusters from the atomic limit, adding one atom at a time. The exchange interaction is, surprisingly, observed to drastically increase compared to that of bulk, and to exhibit irregular oscillations as a function of the interatomic distance. From electronic structure theory we find that the theoretical magnetic moments oscillate with cluster size in exact agreement with experimental data. Unlike the bulk, the oscillation is not caused by the RKKY mechanism. Instead, the inter-atomic exchange is shown to be driven by a competition between wave-function overlap of the 5*d* shell and the on-site exchange interaction, which leads to a competition between ferromagnetic double-exchange and antiferromagnetic super-exchange. This understanding opens up new ways to tune the magnetic properties of rare-earth based magnets with nano-sized building blocks.

**Introduction**

Magnetism is a macroscopic phenomenon that at microscopic level occurs due to exchange interactions, whose typical range, or more simply length scale, is determined by the spatial extent of the quantum mechanical wavefunctions [1]. Confinement of these wavefunctions by for example the presence of a surface leads to many unusual magnetic phenomena [2]. A natural question, in light of these considerations, is what happens in a system smaller than the length scale of the bulk exchange field? Here we investigate Tb clusters as a representative of magnetism of rare-earth clusters, and we try to draw conclusions which apply in general to clusters of rare-earth elements.

The rare-earth metals have similar crystal structures, which arise from electronic structure of the valence shells as the localised 4*f* shell is being filled [3-6]. In spite of this, their magnetic structures vary significantly [7]. This is directly related to the mechanism of the exchange interaction in these materials where the spin-polarized 4*f* wavefunctions of each atom do not overlap but are responsible for a large magnetic moment. In contrast, the *spd* electrons are delocalized and form bands, leading to rather complicated Fermi surfaces [8]. The exchange interaction known as Ruderman-Kittel-Kasuya-Yosida (RKKY) interaction [9-11] between the localised 4*f* moments is mediated by these delocalised electrons. It is long-ranged and results in the oscillatory behaviour of the magnetic coupling as a function of the separation between the atoms [12], described by the electron wave-vectors at the Fermi surface. In a periodic lattice this causes certain frustrations and results in several magnetic phases, from simple ferromagnetism to helical antiferromagnetism and others [7,13-15].

The interatomic exchange parameters for the rare-earth metals are relatively small. For Tb they are of the order of 10-100 μeV [16]. Nevertheless, because of the rather long range of this interaction, up to 7-8 Å, a contribution by many sites (66 within a radius of 8 Å) must be summed up to account for the total strength of the effective exchange interaction (the local Weiss field). In the case of bulk Tb, this Weiss field becomes $H_{ex}$ = 8.3 meV/$\mu_B^2$, resulting in a critical temperature above 200 K.

When the system size is reduced, electronic wavefunctions mediating the exchange are constrained by the boundaries. Because of this, surfaces often show magnetic properties drastically different from their bulk counterparts [2,17]. For example, the magnetocrystalline anisotropy energy (MAE) typically increases in low-dimensional systems. In bulk Tb, the MAE is around 1 meV per atom, whereas in the clusters studied here, both experiments and theory give values of MAE of the order of 10 meV per atom (see Supplementary information for details). The situation is expected to become extreme if the size of the whole system becomes comparable to or smaller than the length scale of the bulk exchange interaction. Such a situation is unthinkable in the case of the direct exchange, which by nature is short-ranged, or even the short-ranged super-exchange and double-exchange. In fact, the ranges of these three types of exchange interactions are all limited to the interatomic distances. It is for materials with RKKY exchange that a unique situation occurs, in that the exchange has to be drastically modified for clusters when compared to bulk. For this purpose we have studied experimentally and theoretically the magnetic properties of rare-earth clusters, using Tb clusters as a representative.

**Results**

*Experiment: Stern-Gerlach magnetic deflection*

To study the magnetism of small mass-selected Tb clusters we used a standard laser-ablation cluster source with controllable temperature in combination with Stern-Gerlach magnetic deflection spectrometer (Fig. 1(a) and see 'Methods: Experimental details'), see also an early attempt in Ref. 18. The first striking result of these measurements is that the magnetic moment of the clusters oscillates with the number of atoms, from very large values of about 10 $\mu_B$/atom (note that bulk Tb has an atomic magnetic moment of about 9.3 $\mu_B$/atom [19]) down to 2 $\mu_B$/atom, as shown in Fig. 1(b). Keeping in mind the localized nature of the 4$f$ states for this heavy lanthanide, it is unlikely that the atomic magnetic moments become quenched. Therefore, variations in the exchange interaction, possibly coupled to the geometrical structure of the clusters, must be the reason for these oscillations. The resulting magnetic structure is then expected to be either ferrimagnetic or non-collinear.

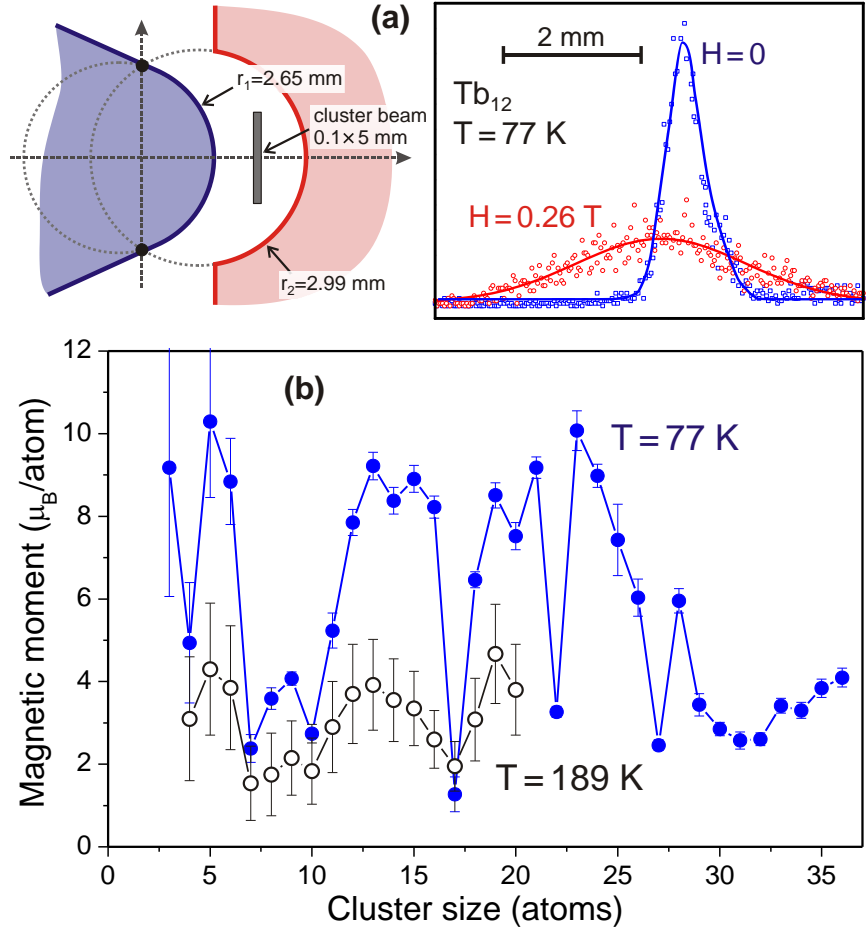

FIG. 1. Magnetic deflection measurement of the magnetic moments of $Tb_N$ clusters: (a) Cluster beam profiles with the deflection magnet (parameters shown in the inset) "on" and "off". While the deflection towards higher magnetic field indicates a sizable magnetic moment, the double-sided deflection profile is a clear sign of a large magnetic anisotropy locking the magnetic moment to the cluster lattice. For the further discussion, see Supplementary information (b) Magnetic moments of $Tb_n$ clusters showing very large oscillations between ferro- and antiferromagnetic configurations in the clusters. While the magnetic moment is strongly reduced at higher temperature, the ferro- / antiferromagnetic variations are preserved.

At T=189 K the magnetization is measured to be reduced to 25-30% of the low temperature values, but nevertheless the clusters are still magnetic. This is also confirmed by the similar size-dependent oscillations in the magnetic moment. Therefore the Curie temperature of the clusters is at least of the same order as in the bulk, where $T_C$ ~230 K [19]. This fact directly indicates that the exchange interaction is drastically different from that of the bulk, since finite size effects normally lead to reduced ordering temperatures, and this is further analysed below.

*Electronic structure: valence stability*

With the Tb cluster magnetic moments established experimentally, theoretical calculations and considerations were employed to understand the origin of the observed variations. However, before the magnetic structure can be investigated, we first need to consider the geometry and electronic structure, since the magnetic structure strongly depends on these quantities. The geometry of selected Tb cluster cations is investigated in Ref. 20, where experimental vibrational spectroscopic data are compared with theoretically calculated ones for Tb cation clusters. In our work, as a starting point, full geometry optimization is performed for neutral Tb clusters up to $Tb_{13}$. These geometries happen to be very similar to those obtained for cations [20], though naturally with some differences in the interatomic distances introduced by the ionization. With this information on the geometry the next step can be taken, namely a consideration of the electronic structure. In particular, the valence stability of Tb clusters should be investigated since a-priori it is unknown whether these clusters are tri- or di-valent. The valence stability refers to the number of valence electrons in a system and thus indirectly to the number of 4*f* electrons. Since, the 4*f* electrons constitute the local magnetic moment and the *spd* electrons (valence electrons) mediate the coupling between them in a rare-earth system, it is absolutely crucial to know the exact number of the electrons involved in these two 'sub-systems'.

For this purpose calculations of the valence stability using the Born-Haber cycle [6,21] were performed, which evaluates the total energy difference between a trivalent and divalent configuration (for all details see 'Methods: Valence stability'). Here trivalent corresponds to the configuration with 3 valence electrons (*spd* electrons) and divalent to the situation with 2. For a Tb atom a trivalent configuration corresponds to a 4*f* shell occupied by 8 electrons, giving rise to S=3, L=3, J=6 and g=3/2 [7], whereas a divalent atom has 9 electrons in the 4*f* shell, with S=5/2, L=5 and J=15/2 and g=4/3. That the preference for either trivalent or divalent for a Tb cluster is not clear from the beginning, can be understood from the fact that the free Tb atom is divalent (as are most of the rare-earth atoms) in contrast to bulk Tb, which forms in a trivalent configuration. Thus it is unknown where the transition from divalent to trivalent takes place when going from atom to bulk, via different sizes of the clusters. Calculating the energetics of the Born-Haber cycle for the valence stability calculations (using density functional theory in generalized general gradient (GGA) limit, for details see 'Methods: Valence stability') we found that all Tb clusters considered in this study are trivalent, at their equilibrium bond distances. Thus all Tb clusters have 8 4*f* electrons and 3 *spd* electrons per atom, and the magnetism of the clusters should be further investigated using this electronic configuration.

## Magnetic structure I: 4f electrons in the core

In order to understand the magnetic behaviour, we employ a theoretical analysis based on *ab-initio* calculations using two complementary theoretical tools, which are extensively described in 'Methods: Theoretical methods and computational details I' and 'Methods: Theoretical methods and computational details II'. The first treats the 4*f* electrons as part of the core states, which means that they are chemically inert within this treatment (see 'Methods: Theoretical methods and computational details I' for more details). In the past this method was used with success for rare-earth bulk systems, and constitutes what is referred to as the Standard Model of the rare-earths. As examples, one may note that the Standard Model was shown to reproduce with good accuracy measured densities, structural stability, bulk modulus and even the interatomic RKKY exchange [4-6,13-15]. The second method, LDA+U, that we have used is discussed in the next paragraph (Magnetic structure II: LDA+U).

Comparing for $Tb_2$-$Tb_7$ the calculated total magnetic moments of this first method (yellow dots) with the experimental values (black squares) in Fig. 2, a few things should be noted: first, the general trend is perfectly reproduced, and second, the absolute values are also very close to the experimental data.

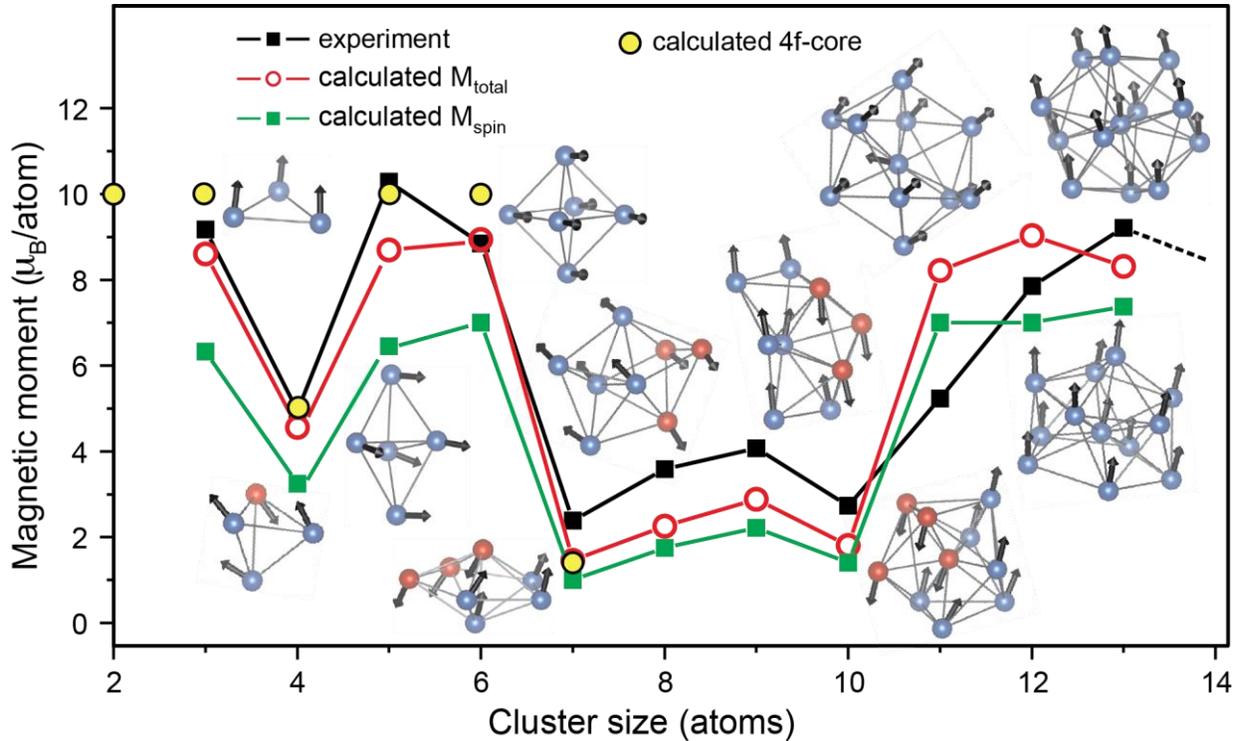

FIG. 2. Calculated geometric structures and magnetic moments of $Tb_n$ clusters. The experimental magnetic moments in $\mu_B$/atom (black squares) are compared with the results of calculations. The yellow

dots show the magnetic moments from 4f-in-the-core calculations. From the LDA+U method, the green filled squares correspond to the spin moment of a collinear calculation without spin-orbit coupling and the red open circles to spin plus orbital moment of a non-collinear calculation with spin-orbit coupling. In addition, the details of cluster structure and magnetic configuration are shown, where blue (red) coloured atoms indicate spin "up" ("down") states.

*Magnetic structure II: LDA+U*

In order to analyse the influence of non-collinearity on the same footing as to study the possibility of a quenching of the 4*f* moment, we performed calculations with the LDA+U method (see 'Methods: Theoretical methods and computational details II') [22,23]. This method also makes use of an effective single electron approximation. However, the 4*f* electrons are now allowed to hybridize with the *spd* electrons. Further, the onsite Coulomb interaction between the 4*f* electrons is treated on a mean field level. It is important to note that non-collinearity may originate from geometrical frustration as well as the Dzialoshinskii-Moriya interaction and single-ion magnetic anisotropy. Therefore, in order to capture all mechanisms on equal footing, we also introduced spin-orbit coupling within the LDA+U calculations.

Comparing the total calculated magnetic moments (red open circles) with the experimental values (black squares) in Fig. 2, a few things should be noted: first, the general trend is also on this level of approximation perfectly reproduced, and the absolute values of the LDA+U calculations are very close to the experimental data. Second, the orbital part of the magnetic moments is somewhat underestimated, which occurs often in calculations based on density functional theory employing LDA, GGA and LDA+U approximations [24-26]. This underestimation can be seen from Fig. 2 by comparing the magnetic moment of a collinear cluster (e.g. $Tb_3$, $Tb_5$ and $Tb_6$) from a calculation where the spin-orbit coupling is neglected (green squares) with the moments obtained from a calculation which includes spin-orbit coupling (red open circles). In this way an orbital moment of approximately 2 $\mu_B$/atom may be deduced, while from Hund's rules a value of 3 $\mu_B$ is expected. Although a small quenching of the orbital moment might be expected, a value of 2 $\mu_B$/atom is a reduction larger than observed experimentally for most Tb systems, and reflects most likely the approximate nature of the LDA+U method when applied to rare-earths [25]. Third, the calculated atomic moments, in particular the 4*f* projected moments, are independent of the particular cluster structure and the spin component of the 4*f* moments is close to 6 $\mu_B$/atom.

Another source that could lead to a discrepancy between the experimental and theoretical values is the temperature. The experimental values of Fig. 2 are obtained at 77K, while the theoretical values correspond to the ground state, i.e. 0K. However, in order to obtain the ground state (or global minimum), the energy landscape has been analyzed in great detail by starting from many different spin densities. From this analysis it could be observed that the energy differences between the global and local minima are substantial with respect to the 77K. Therefore, the influence of the difference in temperature between experiment and theory can be neglected.

Hence, we conclude that the observed non-monotonic trend of the cluster magnetism (Fig. 2) is due to the varying degree of antiferromagnetic/ferromagnetic couplings in addition to a certain degree of non-collinear couplings among essentially unquenched Tb moments. In Fig. 2 we note that several of the clusters are ferrimagnetic with essentially collinear ordering and some clusters are ferromagnetic, also with essentially collinear couplings ($Tb_3$, $Tb_5$, $Tb_6$, $Tb_{12}$ and $Tb_{13}$). One cluster stands out in this regard, namely $Tb_{13}$ which has all moments essentially ferromagnetic, except that of the centre atom which has a large degree of non-collinearity. We will return to the reason for such behaviour of the exchange couplings below. As noted, the spin configuration of the $Tb_{13}$ cluster is practically collinear (except the centre atom). This is confirmed by the experiments, but is in strong contrast with an earlier prediction of a fan-like spin structure for a 13-atom Gd rare-earth cluster [27, 28], where bulk-like behaviour of the exchange was simply projected onto the cluster structure using a simple Heisenberg exchange model [28]. On the other hand, recent DFT calculations of $Gd_{13}$ [29] showed ferromagnetic structure somewhat similar to our results.

Although the calculated magnetic moments of the LDA+U method match the experimental ones very well, there is a problem involved in these calculations. Namely, the 4$f$ occupation is non-integer (with a deviation of about 0.2 electrons), while we tested, using the valence stability calculations, that it should be integer. In order to analyse this further, next we calculate the exchange interaction in the clusters starting with a dimer, and compare the results of the calculations with the 4$f$ electrons in the core with LDA+U results.

*Exchange mechanism*

The calculated total magnetic moments of both the 4$f$ electron in the core and LDA+U method are in excellent agreement with experiment, indicating the correct guess of the internal magnetic structure of each of the investigated clusters. However, the physical mechanisms that determine the couplings between these moments have to be elucidated. In order to obtain an understanding of these couplings we study in more detail the exchange interaction of a Tb dimer as function of interatomic distance. More precisely we calculate the total energy difference between a ferromagnetic and an antiferromagnetic Tb dimer. In Table 1 the first column contains the interatomic distance in Ångström, the second and third columns show the energy difference between a ferromagnetic and antiferromagnetic configuration, ΔE, obtained by, respectively, localizing the 4*f* electron in the core and localizing the 4*f* electrons via the LDA+U method. Here a positive ΔE means that the ferromagnetic configuration is favoured. Thus from Table 1 it is clear that the results of the two methods differ drastically. The former method predicts a ferromagnetic configuration to be favourable for all interatomic distances, whereas the latter predicts a behaviour where the coupling fluctuates between antiferromagnetism and ferromagnetism, depending on distance.

TABLE 1: The first column contains the interatomic distance in Å of the Tb dimer. The second and third column correspond to the total energy difference *ΔE* (in eV) between a ferromagnetic and antiferromagnetic configuration for respectively the 4*f* electron in the core and the LDA+U method. The fourth and fifth columns contain the 4*f* occupation in the LDA+U calculation, the first corresponding to an antiferromagnetic configuration and the second to the ferromagnetic one.

| Interatomic distance (Å) | ΔE (eV) *4f electron in core* | ΔE (eV) *LDA+U* | 4*f* occupation *LDA+U* | |
|---|---|---|---|---|
| | | | AFM | FM |
| 2.5 | 1.05 | -0.82 | 8.2 | 8.3 |
| 3.0 | 1.11 | 0.97 | 8.6 | 8.4 |
| 3.5 | 0.54 | -0.29 | 8.8 | 8.5 |
| 4.0 | 0.49 | -0.04 | 8.6 | 8.03 |
| 4.5 | 0.32 | -0.35 | 8.7 | 8.1 |

The calculated 4*f* occupation is listed in column 4 of Table 1 for the LDA+U method, where the first number in this column corresponds to the 4*f* occupation of the antiferromagnetic configuration and the second to the ferromagnetic configuration. Clearly the 4*f* occupancy is non-integer, which is surprising in light of the integer occupation expected for fully localized electron systems and the Standard Model of the rare-earths. By using the Born-Haber cycle (for details see 'Methods: Valence stability') we find that for 2.5, 3.0 and 3.5 Å bonding distance the 4*f* occupation should be 8 (trivalent), while for 4.0 and 4.5 Å it should be 9 (divalent). This shows that the transition from divalency to trivalency occurs within the diatomic molecule at a bond distance larger than the equilibrium distance. Potential experimental ways to control the interatomic distance to investigate this predicted valence change would be to place Tb atoms on a substrate, where the interatomic distances can be varied by direct placement of the Tb atoms. Actually such an experiment has already been performed with a spin polarized STM [30]. Nevertheless, these considerations show that the occupations given by the LDA+U method in Table 1 are somewhat inaccurate, with a resulting uncertainty in the calculated electronic structure properties (magnetic moments, exchange coupling etc.), and we conclude that the 4$f$ as core results are more accurate than the results of the LDA+U method in Table 1. In light of this it is surprising that the trend in the magnetism of the experimentally investigated Tb clusters (Fig. 2) is reproduced quite accurately with this level of theory. We will discuss this point further in the concluding section.

We now move on to discuss possible exchange mechanisms that can explain the results of Fig. 2 and Table 1. More precisely, only the origin of the isotropic exchange is investigated here. Although it would be very interesting to also address the physical origin of the non-collinearity of the magnetic structure, i.e. Dzialoshinskii-Moriya interaction, geometrical frustration or single-ion anisotropy, it is out of the scope of this work. Furthermore, it is the isotropic exchange interaction, which plays the major role in determining the magnetic structure. We start by noting that the radial extent of the atomic 4$f$ wave functions prohibits a direct coupling to be responsible for the exchange, since the overlap is too small. Instead, an indirect coupling via the *spd* valence states must be responsible. However, in the case of clusters it is rather inappropriate to talk directly about RKKY exchange, since a Fermi surface and Bloch states are not present. Therefore, it is better to speak more generally of indirect exchange. This indirect coupling could be mediated either by a local (intra-atomic) exchange interaction, or by mixing (hybridization) between the 4$f$ and *spd* valence states [14,15]. For rare-earth bulk systems it is well known that the former effect dominates [7]. From our calculations we find that the 5$d$ states give the largest contribution to the exchange. Thus, to understand the microscopic origin of the exchange one has to determine when a localized 5$d$ state prefers to hop to a site with a local 4$f$ moment parallel or anti-parallel to the local 4$f$ moment at its own site.

One may speculate that the exchange mechanism described by Alexander and Anderson [31] could take place in these clusters. According to this theory, ferromagnetic coupling in a bulk material is favoured when the Fermi level is at a peak in the 5$d$ partial density of states and for antiferromagnetic coupling the Fermi level is between two peaks. These pictures exactly correspond to ferromagnetic 'double-exchange' and antiferromagnetic 'super-exchange', respectively. Note that this picture concerns a general physical picture and is not limited to some model approximations, e.g. mean-field treatment of the single band Hubbard model [32,33].

In Ref. 31 the situation of two impurities in an electron gas is considered. The interaction of the impurity with the electron gas leads to a broadening of the impurity peak. Therefore one could consider the $5d$ peak positions with respect to the Fermi level as a fingerprint for the exchange mechanism, e.g. double or super-exchange. However, in case of a cluster we are dealing with discrete energy levels rather than broadened peaks. Therefore we should reinterpret the considered picture in the following way: when the spin sign of the $5d$ partial density of states is the same below and above the chemical potential, ferromagnetic double-exchange is preferred, while for opposite spin sign antiferromagnetic super-exchange is favoured. Note that this interpretation is equivalent to the original one of Alexander and Anderson [31].

For convenience a schematic picture of the double-exchange and super-exchange mechanism together with their corresponding $5d$ partial density of states is depicted in Fig. 3. From this picture it is clear that for the double-exchange situation an antiferromagnetic configuration is unfavourable due to the onsite exchange interaction, e.g. the onsite exchange between the $4f$ and $5d$ electrons, which opposes the hopping. In case of super-exchange the ferromagnetic configuration is unfavoured due to the absence of hopping possibilities with respect to the antiferromagnetic configuration.

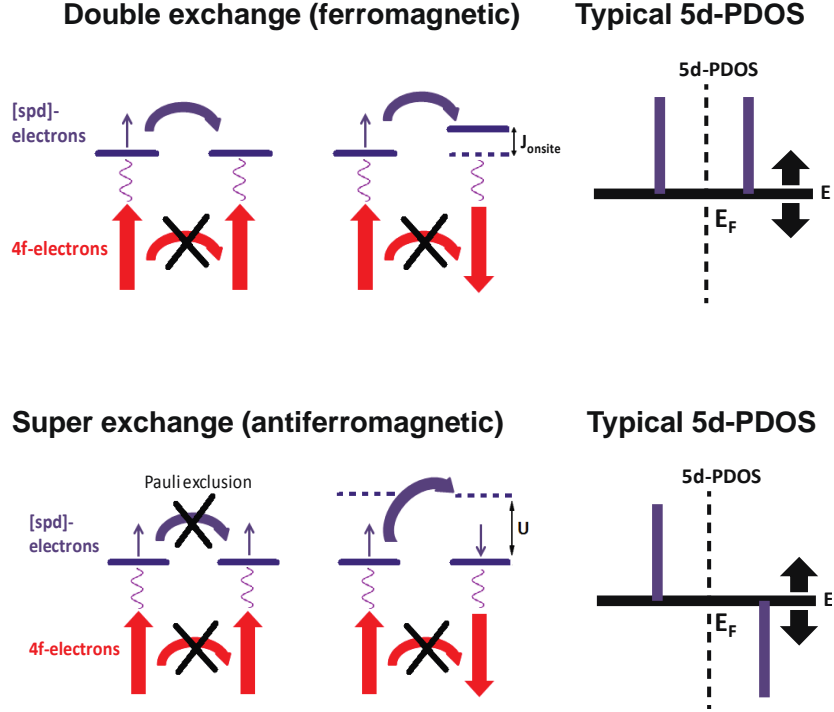

FIG. 3. A schematic picture of the double and super-exchange mechanisms together with their corresponding 5d partial density of states (5d-PDOS). The red arrows correspond to the localized magnetic moments constituted by the 4f electrons, while the small purple arrow corresponds to an *spd* electron.

Thus for the double-exchange situation we would have, without hopping, two degenerate spin up levels at the Fermi level ($E_{Fermi}$ in Fig. 3). With hopping these two levels split up, which results in the typical 5d partial density of states depicted in Fig. 3. For the super-exchange situation (without hopping) a spin up and down level separated by the onsite Coulomb repulsion (indicated by $U$ in Fig. 3) is obtained in the 5d partial density of states. Since this onsite Coulomb repulsion is much larger than the hopping, the hopping will only result in a small shift of these levels. This means that the overall picture of a spin up and down level separated by the onsite Coulomb repulsion in the 5d partial density of states remains the same.

In order to test if the mechanism of Ref. 31 can be applied to the clusters investigated here, we used the method of treating the 4f electrons as core states to calculate the 5d partial density of states as function of inter atomic distance for $Ce_2$, $Pr_2$ and $Tb_2$, and compared it with the total energy difference between a ferromagnetic and antiferromagnetic calculation. Note that for $Ce_2$ and Pr$_2$ this means that respectively 1 and 2 4$f$ electrons are treated as core states of each atom, whereas for Tb 8 4$f$ electrons were treated as core. For Tb dimers we found that a ferromagnetic configuration is favoured for all interatomic distances (Table 1). Therefore in the 5$d$ partial density of states the same spin sign is expected below and above the chemical potential. This is confirmed by Fig. 4(a). Note that the discrete peaks are broadened, which is done for clarity only. For Ce$_2$ we found that for an interatomic distance of 2.5 and 3.0 Å the antiferromagnetic configuration is favoured, while for 3.5 Å ferromagnetism is stable. Thus for 2.5 and 3.0 Å an opposite spin sign above and below the chemical potential is expected in the 5$d$ partial density of states, while for 3.5 Å it should be of the same sign. Fig. 4(b) confirms this picture. For Pr$_2$ an antiferromagnetic configuration is preferred for an interatomic distance of 2.5 Å, while for 3.0 and 3.5 Å it is ferromagnetic. Again this is also what would be predicted from the 5$d$ partial density of states of Fig. 4(c).

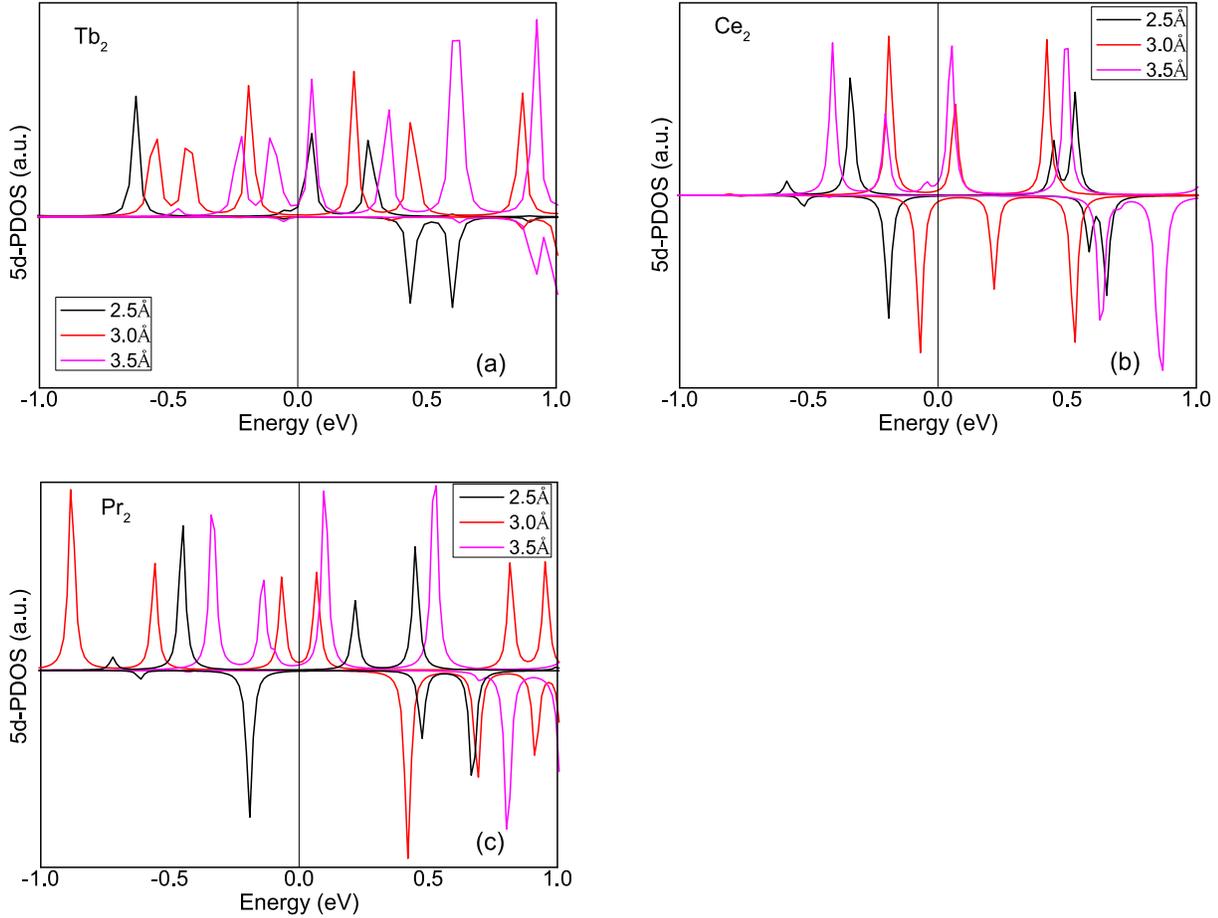

FIG. 4. The 5$d$ partial density of states (5d-PDOS) for an interatomic distance of 2.5Å (black line), 3.0Å (red line) and 3.5Å (pink line) of (a) Tb$_2$, (b) Ce$_2$ and (c) Pr$_2$.

Thus when treating the 4*f* electron as part of the core, the 5*d* partial density of states is consistent with the calculation of the sign of the exchange coupling, e.g. it explains the exchange mechanism. Hence, treating the 4*f* electron as part of the core seems to correctly predict the total magnetic moment, the sign of the exchange coupling, and enables an interpretation of the mechanisms behind the exchange interaction. More precisely the exchange of these clusters appears to be determined by a competition between ferromagnetic double-exchange and antiferromagnetic super-exchange. As a fingerprint of these exchange mechanisms the 5*d* partial density of states can be used. An opposite spin sign in the 5*d* partial density of states above and below the chemical potential signals a super-exchange situation whereas in the case of an equivalent spin sign it is double-exchange.

In principle the same method as described above can be applied to larger clusters in order to obtain the sign of the coupling (antiferromagnetic or ferromagnetic) between two arbitrary sites. However, the features in the 5*d* partial density of states in this case happen to be less pronounced, because a site can have couplings of different sign with different neighbours, and for this reason we will not pursue this analysis further.

We also calculated the exchange couplings for the $Tb_4$, $Tb_6$, $Tb_7$ and $Tb_{13}$ clusters with both the 4*f* electrons treated as part of the core and the LDA+U method. The exchange couplings were obtained by first calculating the total energy differences when one of the atomic magnetic moments of the cluster was reversed, for all sites. Then we mapped the results onto a simple Heisenberg model. The resulting exchange constants are shown in Fig. 5(b) and (c) for respectively the LDA+U method and the 4*f* electron in the core method, together with the calculated geometry of the $Tb_4$, $Tb_6$, $Tb_7$ and $Tb_{13}$ clusters (Fig. 5(a)). In Fig. 5(a) the blue coloured atoms refer to atoms with the local magnetic moments pointing in the same direction, while the red ones indicate atoms with moments pointing in the opposite direction of the blues. Further, the exchange parameters are numbered in such a way that an increasing label of the J's corresponds to an increasing Tb-Tb distance. Thus *$J_1$* corresponds to a Tb-Tb pair with smaller interatomic distance than *$J_2$*. There should be mentioned that for the $Tb_4$ geometry both the LDA+U and 4*f* electron in the core method found a distorted tetrahedron. However, the distortions differ between these methods. In the LDA+U method it is an equilateral triangle with one atom placed some distance above exactly the middle of this equilateral triangle, which leads to two different exchange parameters. While for the 4*f* electron in the core method an isosceles triangle with one atom above this triangle is obtained, resulting in four different exchange parameters. It is the $Tb_4$ geometry obtained with the 4*f* electron in the core method that is depicted in Fig. 5(a). For $Tb_6$ and $Tb_7$ the obtained symmetry of the cluster's geometry is the same for both methods. Although the bond distances differ a little bit. For $Tb_{13}$ the geometry optimized structure of the LDA+U method was used for the exchange parameter calculation with the 4*f* electron in the core method. This was done out of computational reasons. The first thing to note from the exchange constants calculations presented in Fig. 5(b) and (c) is that the cluster's exchange constants are considerably larger than those of the bulk. This is consistent with the relatively high ordering temperature experimentally observed for these clusters. We also note that the exchange energy is typically at least an order of magnitude larger than the MAE (see Supplementary information). For example, for the $Tb_{13}$ cluster, assuming a nearest neighbour exchange interaction parameter *J* of 1.5 meV and 12 nearest neighbours, the total exchange interaction energy at the central Tb atom becomes about 0.8 eV, i.e. almost two orders of magnitude larger than the MAE. By comparing the cluster's exchange constants between the two methods, there can be observed that both the sign and magnitude of the exchange constants can be different. For example for Tb4 the magnitude is quite different although the sign is here the same. That the sign is the same means here that the coupling between blue atom in Fig. 5(a) has a positive sign, while the coupling between a blue and red atom has a negative sign. For $Tb_7$ and $Tb_{13}$ the two methods do not predict for all exchange constants the same sign. In the case of $Tb_{13}$ this could be caused by the usage of the LDA+U optimized geometry for the 4*f* electron in the core method instead of performing an additional geometry optimization.

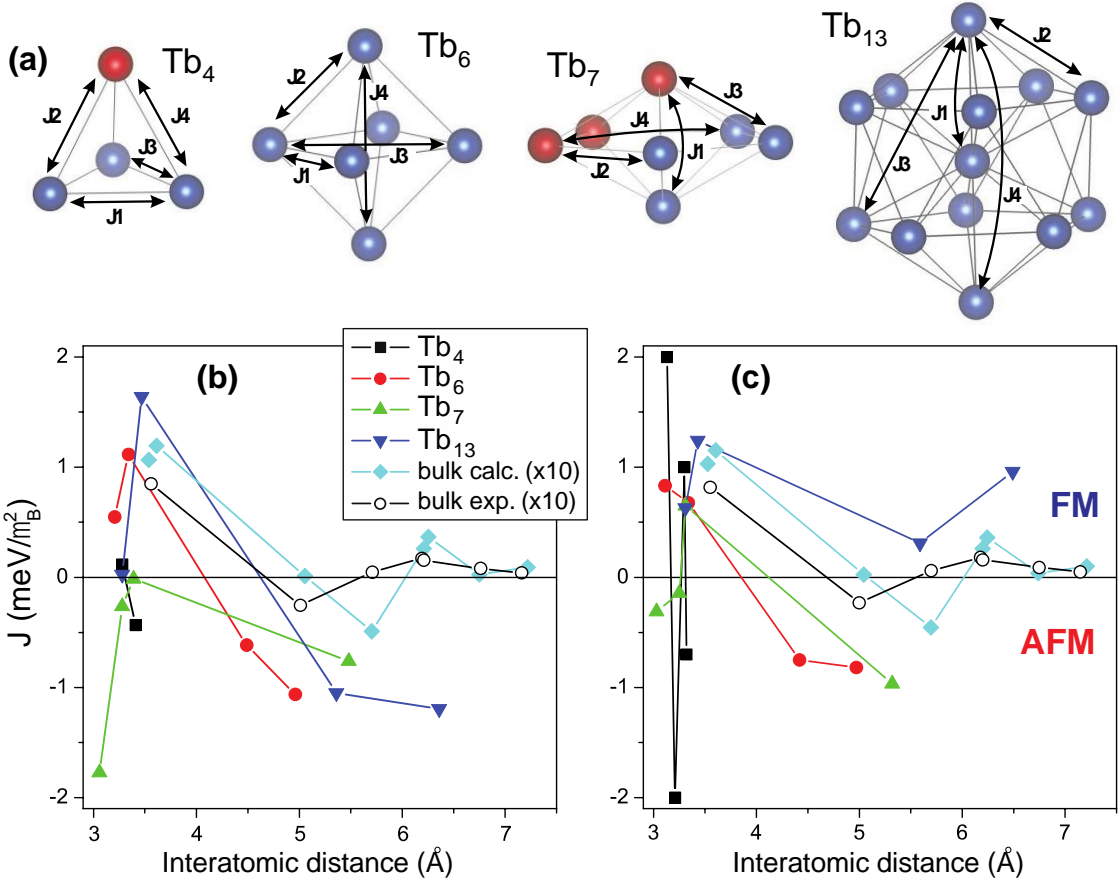

FIG. 5 Exchange interaction parameters in Tb$_n$ clusters and in the bulk: (a) Calculated geometries of Tb$_4$, Tb$_6$, Tb$_7$ and Tb$_{13}$ clusters. The bond distances and interatomic exchange interaction parameters J's are indicated. The blue colour refers to sites with moments pointing in the same direction, while the red ones have moments pointing in the opposite direction of the blue ones. (b) Calculated exchange constants as a function of Tb-Tb separation for the clusters shown in (a).

**Conclusion**

To conclude, we report on an experimental, non-monotonic behaviour of magnetic moments of Tb clusters as a function of size, a variation that is very well reproduced by *ab-initio* density functional calculations, in particular when treating the 4*f* electron in the core and from the LDA+U method. However, we find that the LDA+U results should be interpreted with great care, since non-integer 4*f* occupations are involved leading in some cases to erroneous exchange couplings, e.g. for the Tb dimer. These non-integer 4*f* occupations are in disagreement with the Standard Model and the valence stability calculations with the Born-Haber cycle. Despite this shortcoming, we find that the observation of high and low moment clusters is traced back in both the LDA+U method and the calculation that treats the 4*f* electrons as part of the core, to the presence of oscillatory ferromagnetic and antiferromagnetic exchange interaction between Tb moments as a function of the interatomic distance, an oscillation which originates not from the bulk-like RKKY interaction, but due to an indirect exchange mechanism sensitive to finite size effects. These exchange interactions have a length scale shorter than that of the bulk rare-earths and a significantly larger strength, and are shown to originate from the spatial overlap of the valence band states. The enhanced strength of the cluster exchange interactions, compared to the weaker bulk values, is probably a reason for why the trend of the clusters is captured also by the less accurate LDA+U method. The identification of new exchange mechanisms and the potential for significantly larger interatomic exchange interactions, in combination with unique behaviour of the magnetic anisotropy energy, points to the possibility of designing new rare-earth based magnets with drastically different properties compared to the bulk by utilizing finite size effects and quantum confinement.

**Methods**

*Experimental details*

A beam of cold Tb clusters is produced in a laser vaporization cluster source [34,35] consisting of a cryogenically cooled chamber (0.5 cm$^3$) with an exit nozzle (1 mm diameter). A pulse of cold He gas is injected into the chamber as a 10 ns pulse from a doubled Nd:YAG laser (532 nm, 30 mJ / pulse) vaporizes a minute amount of metal from a metal rod of 1.5 mm diameter in the chamber. The gas has been cooled in a pre-chamber of the source for 40 ms. The metal vapor condenses into clusters. The clusters dwell in the chamber for about 1 ms after which the thermalized clusters are ejected into a high vacuum chamber. The source temperature (15 - 200 K) determines the population of the energy levels of the clusters in the beam. The resulting cluster beam is collimated and passes through a gradient magnet of a standard Rabi two-wire configuration (*B*<2 T, *dB/dz* < 200 T/cm; magnet length $L_{mag}$ =12.5 cm) situated 1 m from the source. The clusters then enter a position sensitive time-of-flight (TOF) mass spectrometer [34,35] placed $L_{TOF}$ = 1 m downstream from the magnet which simultaneously measures their deflections *d* and their masses *m*. This is achieved by a defocusing of the mass-spectrometer, where the linearity of the deflection is assured by an extra electrode [36]. The clusters are ionized by an ArF excimer laser (193 nm / 6.45 eV, 10 ns pulses, 10 mJ / pulse).

The resolution of the mass-spectrometer (>10000) is sufficient to separate the signal from a Tb cluster with a single hydrogen attached from that of a pure Tb cluster of the same mass. Thus, any impurity affected clusters are filtered out and not considered. We should add here, that the whole time scale of the experiment is of the order of a millisecond. Thus, it is still a very small fraction of the total number of clusters that gets any impurities.

Cluster velocity $v$ is determined using a chopper in the cluster beam. The period of the chopper and all other distances and delays are known, so by synchronizing the chopper and the gas pulse so that the beam travels through a slit in the middle of the chopper, both the velocity of the beam and the dwell time of the clusters in the source can be calculated.

The exact value of the field and field gradient experienced by the cluster beam depends on the alignment of beam with the magnet. Because this can be measured with low precision, we calibrate the field and field gradient by performing a deflection experiment on the $^2P_{1/2}$ ground state of the aluminum atom, that has a magnetic moment of $\mu_B/3$.

*Interpretation of the deflection profiles*

The interpretation of the deflection profiles is central to the analysis of the response of a cluster to a magnetic or electric field. In general one can categorize two situations that can occur when a particle such as a cluster is deflected by an applied field:

(i) when the moment is fixed to the lattice, that is it has a preferred orientation, the cluster will undergo a nutational motion as it passes through the field and can be described by a rigid rotor deflection [37]; and

(ii) when the moment has no preferred orientation with respect to the lattice, and thus aligns with the applied field.

The differences in the coupling strength between the moment and the lattice will manifest in the deflection profiles seen; in case (i) where the moment is locked, a two-sided broadening of the profile will be seen, and in case (ii) where the moment is free a single-sided deflection will be observed.

In the first case, one finds

$$\langle M \rangle = \frac{2\mu^2 B}{9kT} + \ldots \qquad (1)$$

where only the first term counts for a spherical particle, and this is still approximately valid in the case of a symmetric top.

If the magnetic moments are free, the deflection is approximately described by the Langevin formula:

$$\langle M \rangle = \mu \left[ \coth\left(\frac{\mu B}{kT}\right) - \frac{kT}{\mu B} \right] \approx \frac{\mu^2 B}{3kT}, \qquad (2)$$

the latter true in the case of small fields or high temperatures.

In the case of Tb clusters, a two-sided broadening is clearly visible in the deflection profiles (see Fig. 1(a)), therefore the rigid-rotor model was assumed.

*Valence stability*

For the bulk it is well established that all rare-earth elements have a trivalent electronic configuration, with the exception of the α-phase of Ce, Eu and Yb. Eu and Yb, in particular, are divalent, since this configuration provides a half-filled or filled 4$f$ shell. On the other hand the rare-earth atoms have an electronic configuration that could be referred to as 'divalent'. The interesting question is what happens in the regime between these two limits, e.g. the regime of clusters. Here the main focus is on the energy difference between a divalent $f^{n+1}[spd]^2$ and trivalent $f^n[spd]^3$ configuration. Unfortunately, this energy difference is not directly accessible in DFT, due to the localized 4$f$ states, which are not properly described by local or semi-local exchange correlation functionals, such as LDA or GGA. These problems can be overcome by means of the Born-Haber cycle [6,21], which is schematized in Fig. 6.

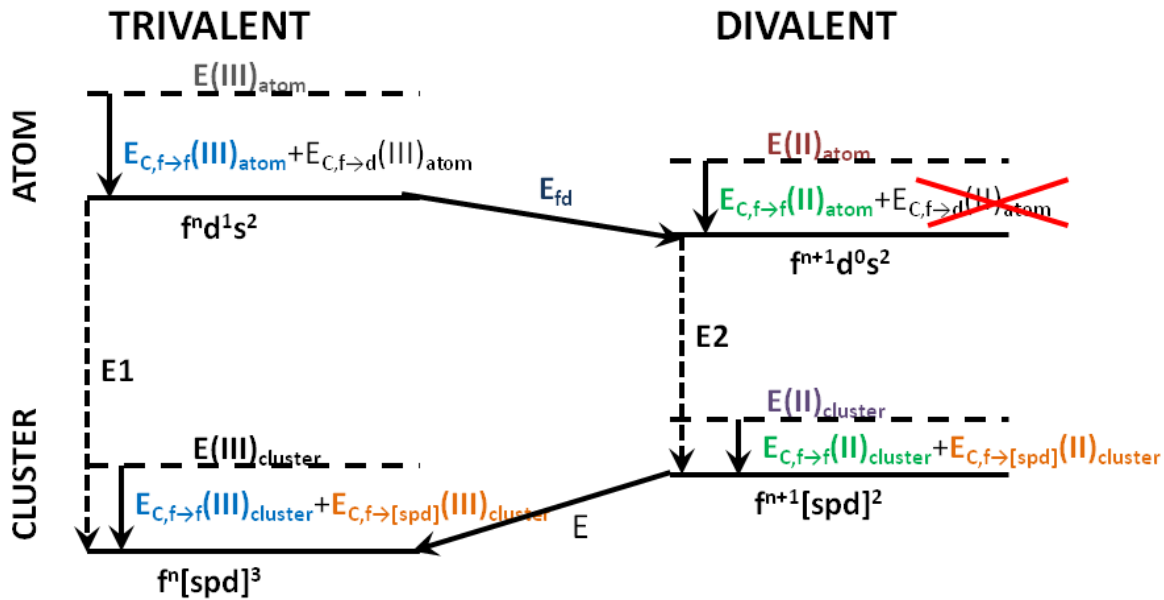

FIG. 6 Schematic picture of the Born-Haber cycle for rare-earth valence stability.

The first assumption of the Born-Haber cycle is that the decoupled states, where the intra 4*f* shell coupling and inter 4*f*-5*d* coupling are neglected, can be accurately calculated within DFT. These states are represented with dashed lines in Fig. 6. The second assumption is that the inter 4*f*-5*d* coupling in the divalent and trivalent clusters is approximately the same (orange formulas in Fig. 6). Thus, now we only need to evaluate the difference in the intra 4*f* shell coupling between the divalent and trivalent clusters. For this purpose we exploit the fact that the 4*f* shells in the isovalent atoms and clusters are approximately the same, due to strong electronic localization. Then, by going around the cycle via the atomic energies we see that the intra 4*f* couplings cancel between the isovalent atoms and clusters (blue and green formulas in Fig. 6). Finally the last terms to evaluate are the atomic *f-d* promotion energy, $E_{fd}$, and the coupling between the 4*f* and 5*d* shell in the atoms, $E_{C,f \to d}(III)_{atom}$. Note that there is no coupling between the 4*f* and 5*d* shell in the divalent atom. Further, the atomic energies can be obtained from experiment. So the final expression for the energy difference between a divalent and trivalent configuration becomes,

$$E = \left[\left(n \cdot E(III)_{atom} - E(III)_{cluster}\right) - \left(n \cdot E(II)_{atom} - E(II)_{cluster}\right)\right] - n \cdot \left(E_{fd} + E_{C,f \to d}(III)_{atom}\right) \quad (3)$$

The method outlined above was used by Delin et al. [6] for the calculation of the valence stability of rare-earth bulk systems. Their results were in agreement with experiment within an error of 0.15eV. In the present study, we used this approach to calculate the valence stability of Tb-clusters for sizes varying from 2 till 8 atoms. All these clusters were found to be trivalent, and with increasing size the trivalent configuration is more and more stabilized.

For the DFT calculations required for the valence stability, the full-potential linear muffin-tin orbital method (FP-LMTO) developed by Wills et al. was used [38]. We used a GGA exchange-correlation functional, in the formulation of Perdew, Burke and Ernzerhof [39]. For the valence states, pseudocore 5*s* and 5*p* basis functions, and the valence 6*s*, 6*p* and 5*d* basis functions, were used. Further, the valence states were treated scalar relativistically (spin-orbit coupling neglected), while the electronic core states (including 4*f* electrons) were treated fully relativistically. For geometry optimization the wrapped polyhedron relaxation method is used [40].

*Theoretical methods and computational details I*

The first method used to calculate the total magnetic moment of Tb clusters is referred to in the main text as the 4*f* electron in the core method. For this type of DFT calculations, the full-potential linear muffin-tin orbital method (FP-LMTO) developed by Wills et al. was used [38]. We used a LDA exchange-correlation functional, in the formulation of Perdew and Wang [40]. For the valence states, pseudocore 5*s* and 5*p* basis functions, and the valence 6*s*, 6*p* and 5*d* basis functions, were used. Further, the valence states were treated scalar relativistically (spin-orbit coupling neglected), while the electronic core states including the 4*f* electrons were treated fully relativistically. For geometry optimization the wrapped polyhedron relaxation method is used [40]. Finally, these calculations were performed for collinear spin structures only, which means that the spin degrees of freedom are only allowed to point up or down.

*Theoretical methods and computational details II*

For the other calculations of magnetic structures, in the main text referred to as LDA+U, the Vienna ab initio simulation package (VASP) was used [42], which is based on density functional theory (DFT). The exchange-correlation functional was considered in the local density approximation (LDA) or the generalized gradient approximation (GGA). These approximations are respectively derived from the homogeneous or nearly homogeneous electron gas, and usually work well for itinerant electrons in solids. When the (valence) electrons acquire a more localized character, e.g. in atoms, molecules or clusters, the strong electron-electron interaction cannot be properly described through LDA or GGA. The errors generated by an approximated exchange-correlation functional are often repaired through methods that go beyond DFT. One of the most successful of these methods is the LDA+U approach, which is used in the present study. First, an explicit two-particle Hubbard-like term is introduced in the Kohn-Sham Hamiltonian to describe the strong Coulomb repulsion between a given set of localized electrons. This term is fully specified by the Hubbard parameter $U$ and the Hund exchange $J$. Then, the resulting Hamiltonian is solved in the mean-field (static) Hartree-Fock approximation. The LDA+U approach is effectively a one-particle theory, but has been proved to be a good compromise between accuracy and computational effort for atomic-like electrons, like the 4*f* states of Tb. Here, we used LDA+U in the rotationally invariant formulation of Dudarev et al. [23]. We redirect the reader to this reference and to [42] for a more complete description of these computational tools.

The Vienna ab initio simulation package (VASP) is a DFT implementation based on a pseudopotential plane wave method and uses periodic boundary conditions. To calculate the electronic structure of an isolated cluster a large cell in a periodic cubic lattice was considered. The size of the unit cell was varied, with edges of 16 Å, 18 Å and 20 Å. The k-mesh was set to contain uniquely the Γ-point. Finally, the cut-off of the plane waves was equal to 400 eV.

The calculations were considered converged for changes of the total energy smaller than $10^{-6}$eV between two consecutive iterations. The geometry was instead considered converged, when the forces on all atoms were smaller than 5meV/ Å.

The exchange-correlation functional was described in the GGA by Perdew-Burke-Ernzerhof in combination with a projector augmented wave method [39]. As mentioned above, LDA+U (note that this is the usual nomenclature and that it is in our case corresponding to GGA+U) calculations were performed in the rotationally invariant formulation of Dudarev et al. [23]. This means that only the difference U-J enters in the total energy functional. The effective U-J value used for the magnetic structure calculations was 3eV, but values up to 5eV were checked to verify that the magnetic structure does not change significantly for a small change of parameters. Finally we should mention that for the magnetic structure of Fig. 2 we performed two types of calculations: (1) A spin polarized calculation without spin orbit coupling and non-collinearity, which means that the spin degrees of freedom are decoupled from the lattice and can only be up or down. (2) A non-collinear spin polarized calculation with spin orbit coupling, meaning that the spins are now coupled to the lattice and are allowed to point in an arbitrary direction. Furthermore, the calculation of the exchange parameters of Fig. 3 was a type (1) calculation and the magnetic anisotropy calculation (see Supplementary information) was of type (2).

**Acknowledgments:** B.S., O.E. and A.D. acknowledge the Swedish Research Council and Carl Tryggers Stiftelse for financial support. Some of the calculations were performed on resources provided by the Swedish National Infrastructure for Computing (SNIC) at the National Supercomputer Center (NSC), the High Performance Computing Center North (HPC2N) and the Uppsala Multidisciplinary Center for Advanced Computational Science (UPPMAX). A.D. also acknowledges financial support from Swedish Royal Academy of Sciences ( KVA) and the Knut and Alice Wallenberg trust (KAW). O.E. acknowledges support from the KAW foundation as well as the ERC (project ASD). Also the Nederlandse Organisatie voor Wetenschappelijk Onderzoek (NWO) is acknowledged for giving us a grant on the LISA supercomputer. SURFsara is acknowledged for the usage of LISA and their support. M.I.K. acknowledges a financial support from European Research Council, Advanced Grant 338957-FEMTO/NANO.

**Author contributions:** A.K. and W.d.H. initially designed and coordinated the experimental part of the project; C.v.D., J.B., and A.K. performed the experiments; S.G., B.S., and L.P. performed the calculations; O.E., M.I.K. and A.K. coordinated work on the paper with contributions from B.S., A.D., I.d.M., B.J., L.P. and discussions with all authors.

**Additional information:** Competing financial interests: The author(s) declare no competing financial interests.